\documentclass[fleqn,usenatbib]{mnras}

\usepackage{newtxtext,newtxmath}
\usepackage[T1]{fontenc}

\DeclareRobustCommand{\VAN}[3]{#2}
\let\VANthebibliography\thebibliography
\def\thebibliography{\DeclareRobustCommand{\VAN}[3]{##3}\VANthebibliography}

\usepackage{graphicx}	% Including figure files
\usepackage{amsmath}	% Advanced maths commands
\usepackage{dirtytalk}
\usepackage{wrapfig}
\usepackage{enumitem}
\usepackage{tablefootnote}
\usepackage{soul}

\defcitealias{A1996}{A1996}
\defcitealias{S2002}{S2002}
\defcitealias{F2019}{F2019}
\defcitealias{D2010}{D2010}
\title[Evolution of hydrogen-deficient binary stars]{Hydrogen-deficient binary stars with magnetic braking}

\author[D.A. Bour et al.]{
David A. Bour,$^{1}$\thanks{E-mail: dab93@ast.cam.ac.uk}
Avishai Gilkis,$^{1}$
Christopher A. Tout$^{1}$
\\
$^{1}$
Institute of Astronomy, The Observatories, Madingley Road, Cambridge CB3 OHA, UK\\
}

\date{Accepted XXX. Received YYY; in original form ZZZ}

\pubyear{\the\year{}}

\begin{document}
\label{firstpage}
\pagerange{\pageref{firstpage}--\pageref{lastpage}}
\maketitle

\begin{abstract}
Hydrogen-deficient binary stars comprise one star which has been stripped of its hydrogen through mass transfer to a binary companion. Observations show that the companion is able to accrete several solar masses without spinning up to critical rotation, and so there must be a mechanism to drain spin angular momentum from the accretor. We test magnetically coupled winds and magnetic star-disc coupling as possible mechanisms and find that, while the disc coupling is negligible, the winds are sufficient to allow the accretor to gain mass without spinning up to critical rotation. However, in order to fully replicate observations, time-dependent scalings of the dynamo-generated magnetic field are needed.
\end{abstract}


%%%%%%%%%%%%%%%%%%%%%%%%%%%%%%%%%%%%%%%%%%%%%%%%%%

%%%%%%%%%%%%%%%%% BODY OF PAPER %%%%%%%%%%%%%%%%%%

\section{Introduction}

Many of the stars in the Universe are contained in multiple-star systems. Most commonly there are binary stars \citep{2012Sci...337..444S, D2013} orbiting a common centre of mass. When the system is close enough, as one of the stars evolves and expands, its outer layers can swell beyond its own Roche Lobe (RL) and so become more gravitationally bound to the other star, leading to Roche Lobe Overflow (RLOF) mass transfer. The characteristics of the mass transfer, such as its rate, timescale, and composition, can strongly affect the evolution of both stars and the system as a whole.\par
Here we investigate hydrogen-deficient binaries, defined as binary star systems in which one of the stars shows very little hydrogen. An evolutionary model of such systems was postulated by \cite{S1983}: the system starts out detached, with both stars evolving independently, and remains detached during the main-sequence (MS) life of the more massive star. As this star evolves to a red giant, it expands and overfills its RL, leading to RLOF. This mass transfer is fast, owing first to the fact that the donor star, or \textit{primary}, is more massive at this point, so that the orbit shrinks under mass transfer, and secondly owing to the expansion of the primary as it becomes a red giant. Most of the mass of the primary is expected to be transferred during this stage so that at some point, the accreting star, or \textit{secondary}, becomes the more massive of the two. The orbit then widens under continued mass transfer and the RL of the primary expands. The primary also regains thermal equilibrium when its expansion to a red giant is completed. These mechanisms cause the mass transfer to slow down. The system at this point is an Algol-type binary \citep{E2006}. Eventually, the orbit widens enough for the two stars to disengage. The primary then evolves on to the helium MS and, following this, into a helium giant, in which a carbon-oxygen (CO) core is surrounded by a fusing helium shell and a small amount of leftover hydrogen which contributes a little to the fusion luminosity. The evolution into a helium giant causes the primary to expand enough that it overfills its RL once more leading to a second mass-transfer stage. This second stage is less intense because of the large initial orbital separation of the stars and the fact that the orbit is continuously widening as mass is transferred. However it is extremely important in the classification of the system because this second mass-transfer stage is where the remainder of the hydrogen-rich envelope is removed from the primary. Owing to the widening of the orbit and the removal of the hydrogen envelope of the primary, the stars then disengage once more, and the primary evolves to a CO white dwarf.\par
A well observed example of a hydrogen-deficient binary is $\upsilon$ Sgr. According to \cite{S1983}, we are currently observing $\upsilon$ Sgr in its second mass-transfer stage. A model atmosphere for $\upsilon$ Sgr, which fits observational data derived by \cite{G2023}, serves as a reference point for the models in this work. Some relevant derived properties of the stars and the orbit are listed in Table \ref{tab:upsag}. Of particular importance is the surface hydrogen mass fraction $X_\mathrm{s}$ of the primary. In this table, $L$ represents a luminosity, $R$ a radius, $M$ a mass, $v_\mathrm{rot}$ a rotational velocity, and $i$ the inclination of the system. The subscript p indicates that a parameter refers to the primary, while a subscript s represents the secondary.\par
\begin{table}
    \renewcommand{\arraystretch}{1.5}
    \centering
    \caption{Spectroscopically derived parameters for $\upsilon$ Sgr}
    \begin{tabular}{|c|c|}
    \hline
        $X_\mathrm{s}$ (primary) &  0.001 $\pm$ 0.5 dex\\
        log$L_\mathrm{p}/\mathrm{L}_\odot$ &  3.67 $\pm$ 0.15\\ 
        log$L_\mathrm{s}/\mathrm{L}_\odot$ &  3.1 $\pm$ 0.2\\ 
        $R_\mathrm{p}/\mathrm{R}_\odot$ &  $28^{+9}_{-7}$\\ 
        $R_\mathrm{s}/\mathrm{R}_\odot$ &  $2.2\pm0.3$\\ 
        $M_\mathrm{p}/\mathrm{M}_\odot$ &  $0.3^{+0.5}_{-0.2}$\\ 
        $M_\mathrm{s}/\mathrm{M}_\odot$ &  $6.8\pm0.8$\\ 
        $v_{\text{rot}, \text{s}}\,\text{sin}(i)/\mathrm{km}\,\mathrm{s}^{-1}$ & 250 $\pm$ 20 \\ 
    \hline
    \end{tabular}\\
    \label{tab:upsag}
    Note: The spectroscopically derived mass of the primary is seemingly impossible for this system from a stellar structure point of view, because low-mass evolved helium stars do not expand into helium giants. However, the evolutionary model favours a mass at the upper end of the spectroscopic range, around $ 0.8\,\mathrm{M}_\odot$, which is still in agreement.
\end{table}
The current orbital period of the system is $138\,\rm{d}$ \citep{K2006}.
\cite{G2023} also infer a starting mass of approximately $5\,\mathrm{M}_\odot$ for the primary, indicating that significant mass transfer has taken place and the secondary has accreted multiple solar masses. However, conservation of angular momentum ought to mean that the secondary should be spun up to critical rotation even after accreting much less (\citealt{P1981}, \citealt{R2021}). \textit{Critical rotation} is defined as the point at which the star is rotating fast enough that the centrifugal force and the gravitational force are equal at the surface. Thus, at critical rotation accretion should cease. The rotational velocity of the secondary is 50 to 90 percent of critical rotation depending on inclination. This was estimated from observations of the circumbinary disc by \cite{N2009}. It indicates that there is some mechanism removing significant angular momentum from the secondary, allowing it to continue rotating subcritically even though several solar masses have been transferred.\par
Our investigation focuses on magnetic effects as possible mechanisms. Specifically we consider a wind coupled to the star's magnetic field and the coupling of the star's magnetic field to its accretion disc. This choice is motivated by the fact that the secondary should have a strong magnetic field owing to its differential rotation (see Section \ref{TSF} for details). The angular momentum carried away by a magnetically coupled wind can be much more significant than that carried away by a normal stellar wind. According to \cite{W1967} and \cite{M1968}, the angular momentum carried away by a magnetic wind is equivalent to that of material corotating out to the Alfvén surface. The Alfvén surface is defined as the surface at which the wind velocity surpasses the Alfvén speed $B/\sqrt{\mu_0 \rho}$. A stronger magnetic field increases the radius of the Alfvén surface for a given wind mass-loss rate. This radius can often be many times that of the star for a strong magnetic field. Consequently, the angular momentum carried away by a magnetic wind can be much larger than that carried away by a wind launched from the star's surface. Magnetic star--disc coupling occurs when a stellar magnetic field interacts with that of an accretion disc around the star. Through this magnetic coupling, angular momentum can be transferred between the star and the disc along field lines \citep{G1978}. The importance of star-disc coupling also depends on the strength of the stellar magnetic field because when the field is stronger more angular momentum is transferred between the star and disc.\par

\section{Input Physics}
We use the Modules for Experiments in Stellar Astrophysics (\textsc{mesa}) code, version 15140 \citep{P2011, P2013, P2015, P2018, P2019, 2023ApJS..265...15J} to evolve our stars. Both components of the binary system have an initial metallicity of $Z=0.014$. The primary has an initial mass of $5.5\, \mathrm{M}_{\odot}$ and the secondary has an initial mass of $2.75\, \mathrm{M}_{\odot}$. \textsc{mesa}'s binary module is used to evolve both stars simultaneously. All simulations are set to end when the surface gravity of the primary exceeds $10^7 \,\mathrm{cm}\,\mathrm{s}^{-2}$, as the primary settles on to the white dwarf cooling track at the end of its life. At this point, the orbital separation of the system has increased significantly owing to the mass transfer, and the two stars have detached from one another. All data analysis is in Python with the \texttt{mesa\_reader} package \citep{BillJosiah2017}. \par
The equation of state in \textsc{mesa} is a combination of OPAL \citep{2002ApJ...576.1064R}, SCVH \citep{1995ApJS...99..713S}, FreeEOS \citep{2012ascl.soft11002I}, HELM \citep{2000ApJS..126..501T}, and PC \citep{2010CoPP...50...82P}. The radiative opacities come primarily from OPAL \citep{1993ApJ...412..752I, 1996ApJ...464..943I}, with opacities at low temperature from \cite{2005ApJ...623..585F} and at high temperature from \cite{1976ApJ...210..440B}. Electron conduction opacities are taken from \cite{2007ApJ...661.1094C}. We use the built-in mesa nuclear reaction network approx21, which uses reaction rates from The Joint Institute for Nuclear Astrophysics REACLIB \citep{2010ApJS..189..240C}. Weak reaction rates are taken from (\citealp*{1985ApJ...293....1F}; \citealp{1994ADNDT..56..231O}; \citealp{2000NuPhA.673..481L}) and screening is implemented according to \cite{1954AuJPh...7..373S}, \cite*{1973ApJ...181..439D}, \cite{1978ApJ...226.1034A} and \cite{1979ApJ...234.1079I}. Thermal neutrino loss rates are implemented following \cite{1996ApJS..102..411I}.\par
The mass-transfer rate between the two stars follows \cite{K1990}. The mass-transfer efficiency $\beta$ is defined so that $|\dot{M}_2|=(1-\beta)|\dot{M}_1|$, where $\dot{M}_2$ is the mass accreted by the secondary and $\dot{M}_1$ is the mass lost by the primary. So $\beta$ is always between 0 and 1. For conservative mass transfer, in which all of the mass lost by the primary is accreted by the secondary, $\beta=0$, while $\beta=1$ means all of the mass lost from the primary is ejected from the system without being accreted. We use a prescription by \cite{G2019}, which increases $\beta$ from zero in three main cases, if the accretion is on the thermal timescale of the secondary, if the secondary fills its own RL, or if the secondary is spinning faster than 95 percent of critical. This prescription replaces \textsc{mesa}'s implicit wind as the star nears critical rotation.\par
We model the specific angular momentum of the accreted mass using a fit to the accretion stream by \cite{dM2013}. When the accretion is direct, the accreted material has specific angular momentum  $\sqrt{1.7GM_\mathrm{a}R_{\text{min}}}$, where $M_\mathrm{a}$ is the mass of the accretor and $R_{\text{min}}$ is the minimum distance of the stream from the accretor, calculated from the fit to the geometry of the stream. When the accretion is through a disc, the material is assumed to accrete with the specific angular momentum of a Keplerian orbit at the surface of the secondary.\par
To model mass loss in stellar winds, we follow the prescriptions of \cite{G2023}. If the star is cool (effective surface temperature $T_{\text{eff}}/\rm{K}\le 10,000$), a relation given by \cite*{dJ1988} is used to calculate the wind mass loss rate. For hot ($T_{\text{eff}}/\rm{K}\ge 11,000$) phases of the evolution the wind prescription is composition-dependent such that, for hydrogen-rich evolutionary stages, $X\ge 0.7$, the wind mass loss is given by \cite{V2021}. For $X\le 0.4$, the relation given by \cite{V2017} is used. For $0.4\le X \le 0.7$, the code interpolates between the two hot-wind prescriptions. Likewise, for $10,000 \le T_{\text{eff}}/\rm{K}\le 11,000 $, the wind mass-loss rate is given by an interpolation of the hot and cool formulae. However, stellar winds are expected to be unimportant to the overall evolution of the system because wind mass loss is dwarfed by RLOF mass transfer. \par
Chemical mixing in \textsc{mesa} is implemented by assigning a diffusion coefficient and effective viscosity to every mixing process. The rotational mixing processes implemented in these models include Eddington-Sweet meridional circulation (\citealt{E1929}, \citealt{S1950}), the Goldreich--Schubert--Fricke instability (\citealt{G1967}, \citealt{F1968}), and the secular and dynamical shear instabilities (\citealt{E1978}, \citealt{P1989}). For explanations of these processes see \cite{H2000}. For convective mixing, stability is determined by the Ledoux criterion \citep{L1947}, which includes composition gradients as well as pressure gradients in the determination of stability. Convection is dealt with through mixing-length theory, as developed by \cite*{H1965}. Convective overshooting, which allows for a small amount of mixing in the areas just outside the boundary of convective zones, is implemented following \cite{He2000}. Thermohaline mixing, where an unstable composition gradient is coupled with a stable temperature gradient, is also included according to \cite*{K1980}. \par
After the primary has transferred most of its hydrogen to the secondary, it begins to transfer helium. This leads to an inverted composition gradient in the outer layers of the secondary owing to uncertainty in the mixing processes in its outer layers. This in turn leads to numerical instability. In order to combat this, the diffusion coefficient in the outer 2 percent of the secondary by mass is artificially increased  to $10^{20} \,\text{cm}^2\,\text{s}^{-1}$, which is slightly higher than that used to model convection, thus effectively mixing the accreted material.\par
Rotation is assumed to be shellular, in that isobaric surfaces rotate as solid bodies. This approximation comes from the fact that mixing by horizontal turbulence is very efficient on isobars, so there is an efficient smoothing of the rotational velocity distribution on each isobar \citep{Z1992}.
\vspace{-.3cm}
\subsection{Magnetic field generation}\label{TSF}
A hydrodynamic dynamo model has long been thought to explain stellar magnetic fields \citep{P1955}. In a differentially-rotating body any field generated by the movement of charged particles is sheared into a toroidal field by the differential rotation. Radial perturbations to the field are smoothed into further toroidal field, increasing the field's amplitude. Eventually, the regeneration of the field is balanced by dissipative effects such as reconnection and an equilibrium is reached. The main application of this mechanism has been in convective dynamos, where the radial perturbations to the field are provided by fluid motion in convective zones (e.g., \citealt{N1984}, \citealt{T1992}). However, the accreting stars of interest here have convective cores and radiative envelopes, so the magnetic field powering the angular momentum losses must be generated in a radiative zone. \citeauthor{S2002}~(\citeyear{S2002}, hereinafter S2002) shows that a dynamo field can be generated in a radiative zone when the perturbations arise from instabilities in the field itself. This Spruit--Tayler dynamo was expanded upon by \citeauthor*{F2019}~(\citeyear{F2019}, hereinafter F2019), leading to the TSF dynamo. In their work, the perturbations to the field are nonlinear and the damping only occurs on small-scales, leading to a larger equilibrium field than was derived by \citetalias{S2002}. This field generation mechanism is the reason we focus our investigation on magnetic effects as responsible for the spin-down, because an accreting star has strong differential rotation in its outer layers and so a strong magnetic field. \citetalias{F2019} give a formula for the equilibrium poloidal magnetic field strength, which we use in all subsequent calculations,
\begin{equation}
    B_{r}=r\Omega\sqrt{4\pi \rho}\left(\displaystyle\frac{q\Omega^5}{N^5}\right)^{1/3},
\end{equation}
where $B_r$ is the poloidal magnetic field strength, $r$ is the spherical radius from the centre of the star, $\Omega$ is the angular velocity of the spherical shell at $r$, $\rho$ the density of the shell, $q$ the dimensionless shear $\mathrm{d}\ln\Omega/\mathrm{d}\ln r$, and $N$ is the Brunt-Väisälä frequency at $r$. \citetalias{F2019} use an effective, thermally-suppressed, Brunt-Väisälä frequency accounting for the efficiency of thermal diffusion outside red giant cores. However, the outer layers of an accreting star, that are the focus of this investigation, do not have this quality and so we use the standard $N$. The dimensionless shear $q$ suffers from numerical zoning in \textsc{mesa}, so that its profile through the star is often quite jagged, so we just assume $q\approx 1$. This means that, over the outer layers of the accreting star, the rotation rate varies on the order of itself. This is not unreasonable when we consider the high levels of shear that result from accretion. Finally, the energy contained in the magnetic field at each meshpoint is limited to less than or equal to 10 percent of the rotational energy at that meshpoint, because the field's energy comes directly from the stellar rotation.
\subsection{Star-disc coupling}
The magnetic torque between star and disc is extremely important in protostellar discs \citep{G1978}. However, protostellar discs are usually more massive than accretion discs, so prior to implementing this torque in our models we would like to verify that the disc is capable, in general, of exerting a torque on the star. For this we use the condition that the disc must have a comparable moment of inertia to the star. The moment of inertia of the star is given by the standard $I_\text{star}\approx0.1MR^2$ for a main-sequence star, where $M$ and $R$ are the mass and radius of the star. That of the disc is given by $I_\text{disc}\approx M_\text{disc}(r_\mathrm{i}^2+r_\mathrm{o}^2)/2$, where $M_\text{disc}$ is the total mass of the disc, and $r_\mathrm{i}$ and $r_\mathrm{o}$ are the inner and outer radii of the disc. The inner radius is given by the magnetospheric truncation radius $r_\mathrm{m}$, defined as the radius out to which the stellar magnetic field disrupts and truncates the disc. It acts as the inner edge of the disc where the stellar magnetic field lines connect and couple. A prescription for $r_\mathrm{m}$ by \cite{T2022} is used. The outer radius is given by the tidal truncation radius $r_\mathrm{t}$ of the disc by the primary, given as $r_\mathrm{t}=0.6a/(1+1/q_\mathrm{m})$ from \cite{W2003}, where $a$ is the separation of the system and $q_\mathrm{m}$ the mass ratio, defined here as the mass of the accretor divided by that of the donor.\par
Finding the mass of the disc requires integrating the surface density profile. We follow the process of \citeauthor{A1996}~(\citeyear{A1996}, hereinafter A1996), to derive the surface density profile of a truncated disc, but in our case we assume the truncation radius of the disc to be the magnetospheric radius. This leads to 
\begin{equation}\label{dens}
\begin{aligned}
    &\nu\Sigma=\frac{\dot{M}}{3\pi}\left(1-\sqrt{\frac{r_m}{r}}\right)+\\
    &\hspace{.3cm}\frac{\mu^2}{9\pi\sqrt{GM}}\left(r^{-7/2}-r_m^{-3}r^{-1/2}-2r_c^{-3/2}r^{-2}+2r_m^{-3/2}r_c^{-3/2}r^{-1/2}\right),
\end{aligned}
\end{equation}
where $\dot{M}$ is the mass-transfer rate incoming to the disc, $r$ is the cylindrical radius from the accretor, $\mu$ is the magnetic moment at the stellar surface given by $\mu=B_\mathrm{s}R^3$ where $B_\mathrm{s}$ is the surface magnetic field, $r_\mathrm{c}$ is the Keplerian corotation radius, and the other variables are already defined. The corotation radius is the radius in the disc at which the Keplerian orbital rotation rate is equal to the stellar rotation rate. So inside the corotation radius the disc rotates more quickly than the star, and outside corotation the disc rotates more slowly than the star. This profile reduces to that derived by \citetalias{A1996} for $r_\mathrm{c}=r_\mathrm{m}$. We use a prescription for the viscosity $\nu$ from \citetalias{A1996}, $\nu=0.3\alpha^{1.05}\Sigma^{0.3}r^{1.25}$, where $\alpha$ is a free parameter taken to be 0.1. Putting this in to Eq. \ref{dens} and numerically integrating, we find the total mass contained in the disc, which when entered into the formula for the moment of inertia of the disc shows that $I_\text{disc}$ is less than $I_\text{star}$ for the entirety of the evolution of the system by approximately two orders of magnitude. As a result, we conclude that the disc cannot exert a significant torque on the star, and we thus neglect the disc torque for the remainder of this investigation.
\subsection{Magnetic winds}\label{magwind}
This leaves the magnetically coupled winds as the only effective mechanism draining angular momentum from the star as it accretes. Our magnetically coupled wind formula follows \citeauthor*{D2010}~(\citeyear{D2010}, hereinafter D2010). They give the angular momentum loss rate in the wind as
\begin{equation}
    \dot{J}_{\text{wind}}=-(\dot{M}_{\text{wind}}^{4n-9}B_\mathrm{s}^8(2GM)^{-2}R^{8n})^{1/(4n-5)}\Omega,
\end{equation}
where $\dot{J}_{\text{wind}}$ is the total angular momentum loss rate of the star, $\dot{M}_{\text{wind}}$ is the wind mass-loss rate from the accretor, $n$ is a factor describing the geometry of the stellar magnetic field, and all other variables are already defined. We follow \citetalias{D2010} in setting $n=3$, which corresponds to a dipolar field. We also use the surface stellar rotation rate for $\Omega$. \par
To calculate $\dot{M}_{\text{wind}}$, we once more follow \citetalias{D2010} in assuming that $\dot{M}_{\text{wind}}=\beta |\dot{M}_{1}|$, where $\beta$ is the mass-transfer efficiency discussed earlier. This leads to a stronger wind if the accretion rate is larger or accretion is less efficient, as it should. For these magnetic loss mechanisms, we set $\beta$ to a minimum of 0.05 rather than its previously imposed minimum of 0 in order to induce a slight non-conservation into the system and allow $\dot{J}_{\text{wind}}$ to be nonzero throughout mass-transfer stages. This also follows the logic of \cite*{D2007} in that, as the secondary spins up, it begins to lose mass from its equator, leading to this slight non-conservation in the mass transfer. The choice of 0.05 follows from the models of \cite*{Da2013} under strong mass transfer. We also tested a minimum $\beta$ of 0.1. The evolutionary difference between the two cases was found to be small, mainly because of the small dependence of $\dot{J}_{\text{wind}}$ on $\beta$.\par
To estimate the surface magnetic field, we take the outer 1 percent of the star by mass and find the maximum poloidal magnetic field given by the TSF dynamo in that region. This relies on the assumption that so close to the stellar surface, this maximum field is able to permeate the small amount of mass above it and control the wind.\par
We also implement magnetic dead zones following \cite{M1987} and \cite*{S2023}. Magnetic dead zones form around a star because not all of its field lines are open. As a result, only a fraction of the particles channelled along the field lines are able to escape the star in a wind. The rest of the particles are trapped in a region inside the closed field lines. This is the dead zone. Our implementation of these dead zones thus reduces the effect of a magnetic wind. With formulae from \cite{M1987} the radial extent of the dead zone at the equator of the star for a dipolar magnetic field can be found for a given set of stellar parameters. We then follow \cite{S2023} to calculate the fraction of the stellar surface that successfully launches a wind using the extent of the dead zone, which is applied as a multiplicative factor to $\dot{J}_{\text{wind}}$. This factor is constrained between 0 and 1. If the factor is 0, it implies that the dead zones extend over the entire star, and thus the wind is completely stifled. If the factor is 1, there is no dead zone, and the wind is launched from the entire star. A plot of this scale factor over the evolution of the system can be found in Appendix \ref{Jdotapp}.
\subsection{Removing angular momentum}
The major challenge with implementing these formulae is that they give only the total angular momentum loss of the star. In these models the star does not rotate as a solid body. \textsc{mesa} models the specific angular momenta of spherical shells so this total $\dot{J}$ needs to be transformed into a specific angular momentum taken away from each layer of the star. To accomplish this, we assume that the angular momentum is only removed from the radiative envelope of the accretor and that the transport of angular momentum in this envelope efficiently smooths out the specific angular momentum profile. This implies that $\dot{j}_\text{layer}\propto r^2$, where $\dot{j}_\text{layer}$ is the specific angular momentum removed from each layer of the star. 
\begin{equation}
    \dot{j}_{\text{layer}}=\frac{1}{\displaystyle\sum_{i} r_i^2\delta m_i}\dot{J}r_{\text{layer}}^2,
\end{equation}
where $\delta m$ is the mass of each shell. These angular momentum losses only apply to the secondary.\par
\section{Results and Discussion}
\begin{figure*}
    \includegraphics[width=\textwidth, alt={The evolutionary tracks of the two stars on a Hertzprung-Russell Diagram are shown here. The tracks are as one would expect, with two mass-transfer stages shown cleary in the track of the primary and a final evolutionary stage of a carbon-oxygen white dwarf.}]{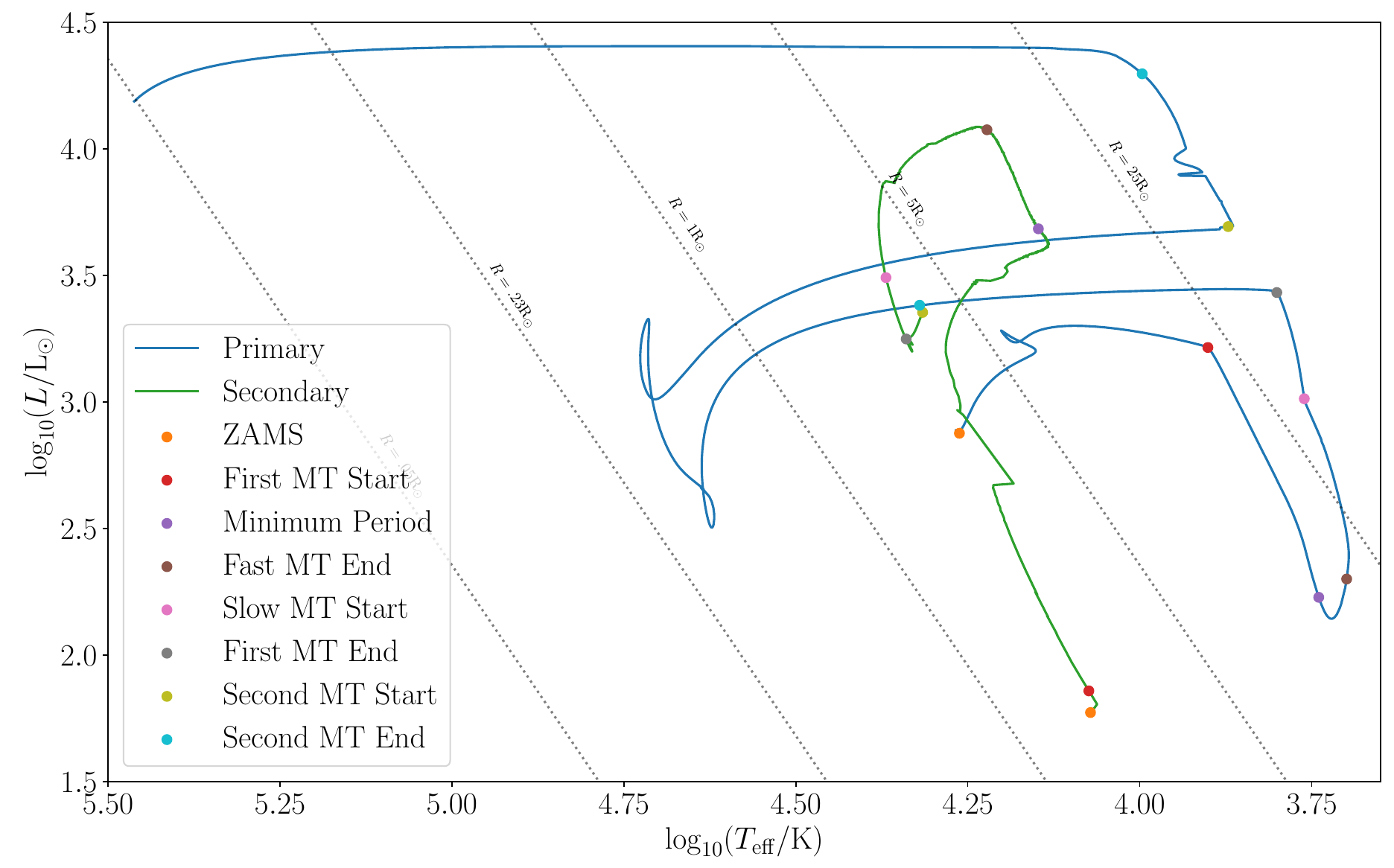}
    \vspace{-.7cm}
    \caption{Stellar evolutionary tracks for a hydrogen-deficient binary system. The blue track represents the primary (donor) star and the green track represents the secondary (accretor) star. The dotted lines are lines of constant radius and the points represent important evolutionary stages. The points are consistent between the primary and the secondary. The beginning and end of all mass-transfer stages are shown, as well as the zero-age main sequence (ZAMS) of both stars and the point at which the period of the system achieves a minimum.}
    \label{HRD}
\end{figure*}
\begin{figure*}
    \centering
    \includegraphics[width=\textwidth, alt={This plot showcases the two mass-transfer stages of the evolution of this system. We see that the first is much more intense, with rates of $10^{-4}\,\mathrm{M}_\odot$, while the second is much more tame, with rates of $10^{-7}\,\mathrm{M}_\odot$}]{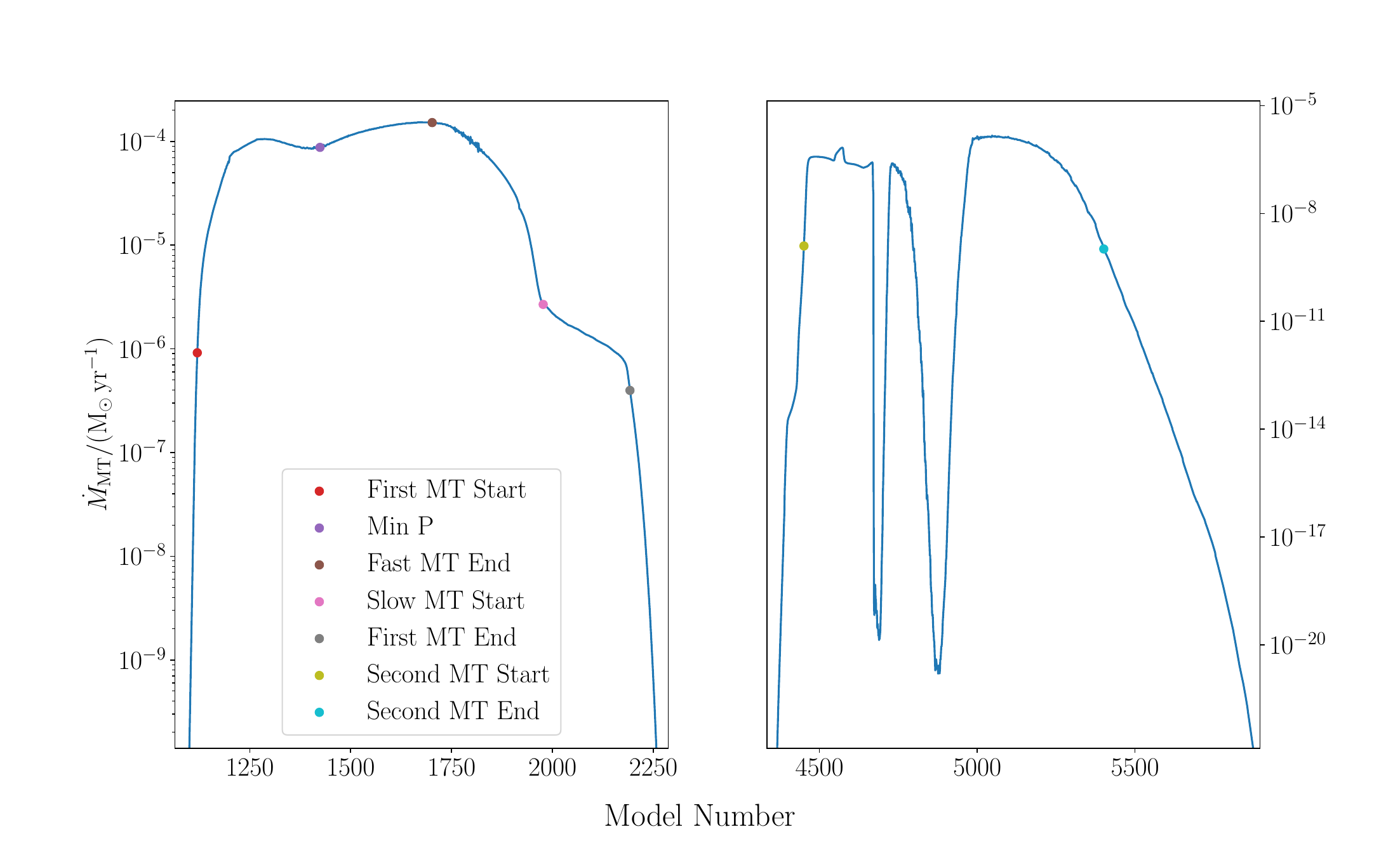}
    \vspace{-.5cm}
    \caption{Mass-transfer rate with respect to model number. The first mass-transfer stage is in the left plot and the second on the right. We can see that the first stage is indeed much more intense and is composed of a fast portion and a slow portion. The second stage has a much smaller mass-transfer rate and because of the large initial separation of the stars and the instability of the primary as it expands to a helium giant, is more erratic than the first.}
    \label{MT}
\end{figure*}
\begin{figure}
    \includegraphics[width=8.5cm, alt={The masses of the two stars as the system evolves are shown. The secondary starts at $2.75\,\mathrm{M}_\odot$ and finishes the first mass-transfer stage at $6.5\,\mathrm{M}_\odot$, before accreting about $0.2\,\mathrm{M}_\odot$ during the second mass-transfer stage. The primary starts at $5.5\,\mathrm{M}_\odot$ and ends at slightly less than $1\,\mathrm{M}_\odot$}]{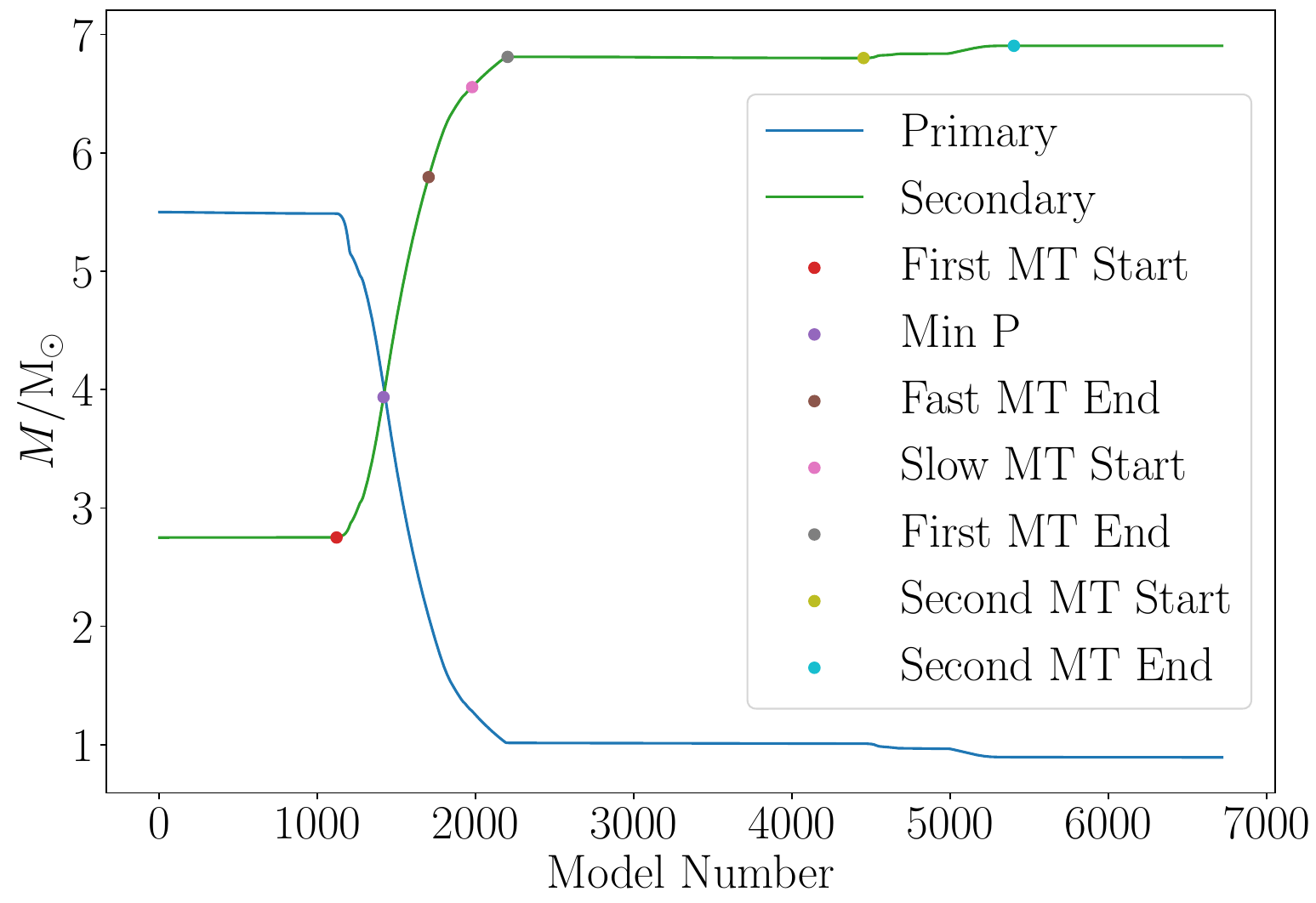}
    \vspace{-.3cm}
    \caption{Masses of each star as the system evolves. We see that the secondary reaches about $6.8\,\mathrm{M}_\odot$ and the primary falls to about $0.8\,\mathrm{M}_\odot$, both of which agree with the observations of this system within 1$\sigma$.}
    \label{Mass}
\end{figure}
\begin{figure}
    \includegraphics[width=8.5cm, alt={The orbital period as the system evolves is shown. The evolution is as expected, with a drop and then increase in the period during the first mass-transfer stage, followed by another, smaller increase in the second, for a final period of approximately 140 days.}]{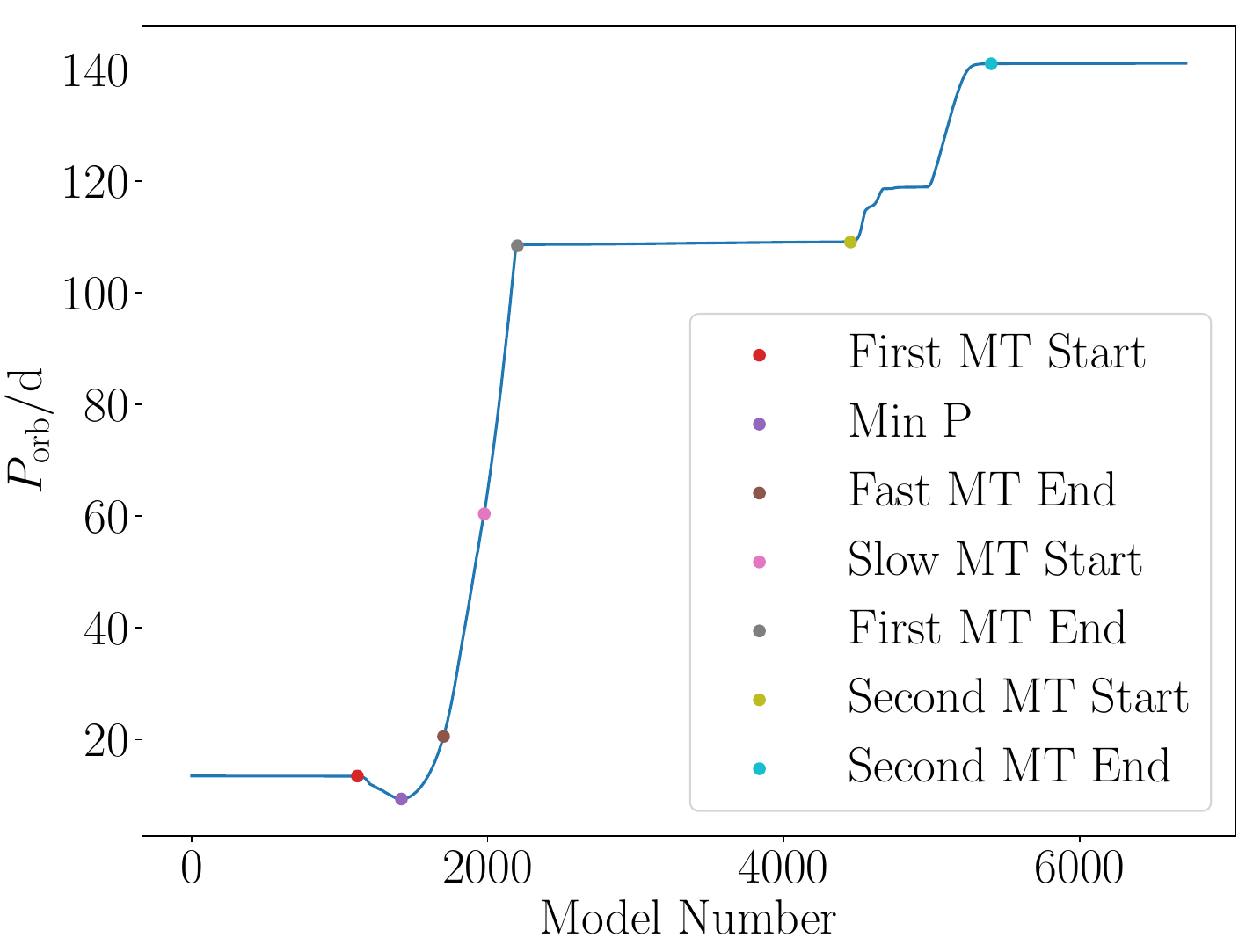}
    \vspace{-.3cm}
    \caption{Orbital period of the system during its evolution. As expected, we see a dip during the first mass-transfer stage followed by an increase once the mass ratio of the stars has inverted, followed by a period of constancy while the stars are disengaged from one another, and another increase corresponding to the second mass-transfer stage. The final period is about $138\,\rm{d}$.}
    \label{Per}
\end{figure}
\begin{figure}
    \includegraphics[width=8.5cm, alt={The surface hydrogen mass fraction $X_\rm{s}$ is shown. The evolution during the first mass-transfer stage drops it to about 0.2. The second mass-transfer stage drops it down to $10^{-3}$, which matches perfectly with observations.}]{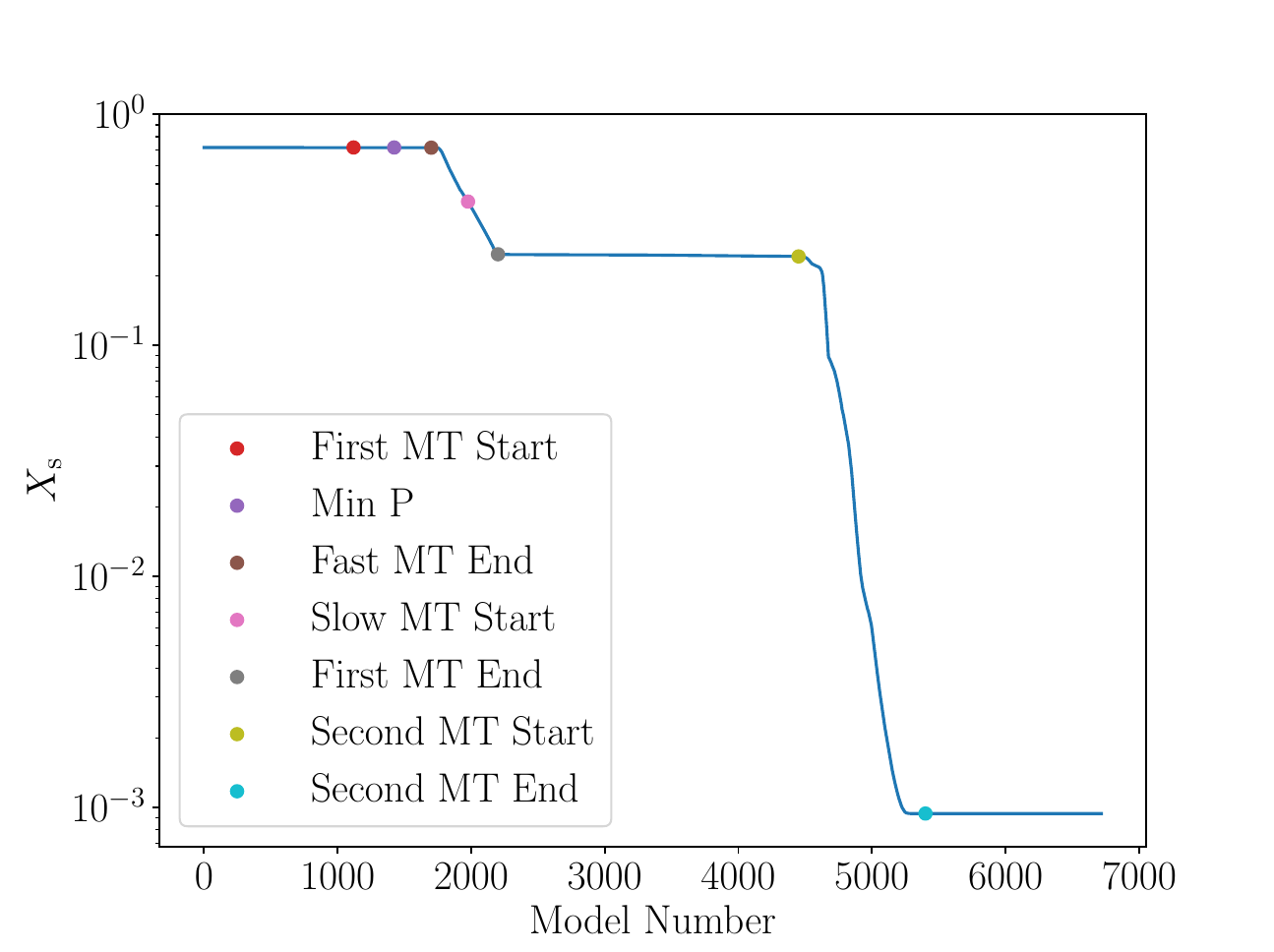}
    \vspace{-.25cm}
    \caption{Surface hydrogen mass fraction of the primary. The first mass-transfer stage ends with the fraction at about $0.2$, while the second drops it to $10^{-3}$, so that the system would be classed as hydrogen-deficient and matches perfectly with the observations of the system.}
    \label{SurfH}
\end{figure}
The evolutionary tracks described here are those that best fit the observations for the current state of $\upsilon$ Sgr as in Table \ref{tab:upsag} as well as the constraints placed on the spin of the accretor during the first mass-transfer stage by observations of Algol-like binaries \citep{vH1990}. The system has initial masses of 5.5~and $2.75\,\rm{M}_\odot$, and an initial period of $13.5\,\mathrm{d}$.\par 
In order to fully match the observations of these systems during both mass-transfer stages, it was necessary to introduce an artificial scale factor of $10^{-3}$ to the magnetic field during the second mass-transfer stage. This is an important requirement of these models. It indicates a gap in our current understanding and modelling of how these fields are generated. One possible reason for the necessity of this scaling is that, while the energy of the magnetic field is constrained by the rotational energy available (see Section \ref{TSF}), the interplay between rotation rate and magnetic field strength is not modelled fully self-consistently. This would have an effect on the models. Another is the dimensionless shear $q$ which, as mentioned in Section \ref{TSF}, was set to 1 implying that $\Omega$ varies on the order of itself in the outer regions of the star. This is an assumption that most likely does not hold for all mass-transfer rates. While it is a better assumption for the first mass-transfer stage because of the much higher rate and much higher differential rotation, the second mass-transfer stage is much less intense. This implies that more of the accreted angular momentum is mixed beyond the accreted layers, so that $q$ is smaller, scaling down the magnetic field. The Brunt-Väisälä frequency in these outer layers is also not well constrained, because the mixing processes in this rapidly rotating, recently accreted material are not well modelled. Finally, as mentioned in Section \ref{magwind}, we assumed the geometry of the field to be dipolar as in \citetalias{D2010}. However, the field in this rapidly rotating recently accreted material is unlikely to be this simple. Observations of highly magnetic white dwarfs show complex field structures, often dominated by higher order multipole components \citep*{2004MNRAS.355L..13T}. These white dwarfs are thought to be a product of binary evolution \citep{2008MNRAS.387..897T}, implying that their frozen-in fields could be indicators of the field structure during the evolution of the star. A different field geometry would alter how the field decays outside the star, thus changing the radius of the Alfvén surface and in turn how much angular momentum the winds are able to remove from the star. It would also change the shape of the magnetic dead zones and thus how they affect the fraction of the stellar surface able to emit a wind. However, the effect of a difference in field structure would most likely be felt more during the first mass-transfer stage. This is because the differential rotation is much stronger in this stage due to the significantly higher mass-transfer rate.\par
It is important to note that, because the magnetic field depends on the stellar rotation rate, this factor does not necessarily scale the strength of the magnetic field by $10^{-3}$ but rather changes where the equilibrium rotation rate at which magnetic braking is balanced by accretion spin-up is placed. This is further discussed in Appendix \ref{bscaleapp}. \par
The Hertzsprung-Russell Diagram (HRD) in Fig. \ref{HRD} shows the evolutionary tracks of both stars in this system, with some of the important points highlighted. The primary evolves normally off the MS until reaching the red point labelled \say{First MT Start}. This is the point at which the primary fills its RL while crossing the Hertzsprung Gap and the first stage of RLOF mass transfer begins. At this point, because the more massive star is transferring mass to the less massive star, the orbit of the system shrinks owing to angular momentum conservation. This in turn implies that the RL of the primary is shrinking. In addition, the primary is still expanding as it continues across the Hertzsprung Gap. This means that the star is expanding as its RL is shrinking, leading to a very high mass-transfer rate (see Fig. \ref{MT}). This is the fast mass-transfer stage. At the point labelled \say{Minimum Period}, enough mass has been transferred that the donor is now the more massive star. This means that the orbit now widens under continued mass transfer. This allows the primary to begin to expand with its RL, so allowing it to begin to climb up the HRD. However, the fast mass transfer continues despite this, because the donor continues to expand on mass transfer. This continues until the orbit has widened enough that the expansion of the donor with mass transfer is mitigated by the expansion of the RL, at which point the mass-transfer rate begins to decrease. This transition period lasts until the pink point labelled \say{Slow MT Start}, where the mass-transfer rate stabilises to about $ 10^{-7}\,\mathrm{M}_\odot\, \mathrm{yr}^{-1}$. The orbit continues to expand during this stage until the primary finally shrinks fully within its own RL and mass transfer ceases. In this stage the majority of mass is transferred. At the end of it the secondary has grown from 2.75 $\mathrm{M}_\odot$ to about $6.7\,\mathrm{M}_\odot$ and the primary has shrunk from $5.5\,\mathrm{M}_\odot$ to about $1\,\mathrm{M}_\odot$. The extended envelope of the primary then collapses but, during this collapse, its mostly helium core becomes hot enough to ignite non-degenerately and it settles on to the helium MS. When this stage ends, the primary is composed of an inert carbon--oxygen (CO) core and concentric helium and hydrogen shells, which begin to fuse as the star expands back across the HRD in its transition to a helium giant. However, before it can complete this expansion, it overfills its RL once more. This second mass-transfer stage is much less intense owing to the large initial separation of the stars and the lack of mass left on the primary to be transferred. However, it is very important in that it removes the last vestiges of the hydrogen envelope of the primary, creating a hydrogen-deficient star. Because the less massive star is transferring mass, the orbit widens under mass transfer and so eventually this mass-transfer stage ends. At this point, the primary is composed of a CO core and an extended helium envelope, which then collapses in a similar manner to the hydrogen envelope after the first mass-transfer stage. However this time the collapse does not generate enough heat for the core to begin fusion and the star becomes a CO white dwarf (WD). This is where we stop modelling.\par
The evolution of the secondary is dictated almost entirely by the mass transferred by the primary. It is still on the MS when the first mass-transfer stage starts and so evolves up the main sequence as it gains mass. The mass transfer at this point is very intense and on a thermal timescale, so that the amount of mass transferred is too much for the secondary to accrete while still remaining on the MS. It instead expands off the MS until slightly before the point of Minimum Period at which the expansion of the star halts, allowing it to begin following a track of nearly constant radius on the HRD. The mass-transfer rate then begins to decrease at the point of Minimum Period and, once the fast mass-transfer stage ends, the secondary can shrink back on to the MS. It begins to evolve off the MS normally before the second mass-transfer stage. It is now a star of about $6.7\,\mathrm{M}_\odot$, so the accreted mass during the second stage does not affect the secondary to a large extent. Only about $0.2\,\mathrm{M}_\odot$ more is transferred.\par
In Fig. \ref{MT} we see the mass-transfer rates in each mass-transfer stage plotted with respect to model number. Model number increases monotonically with age but not uniformly. We use it in place of time to better resolve changes in $\dot{M}_\mathrm{MT}$. The age and model number of the system for each of the labelled points is given in Table \ref{modnum} as a way to translate between the two.\\
\begin{table}
    \renewcommand{\arraystretch}{1.5}
    \centering
    \caption{Model number and age for important evolutionary stages}
    \begin{tabular}{|c|c|c|}
    \hline
        Evolutionary Stage & Model number & System age/($10^7$ Years)\\ \hline
        First MT Start &  1119 & $7.980$\\ 
        Min P &  1423 & $7.983$\\
        Fast MT End &  1701 & $7.985$\\
        Slow MT Start &  1976 & $7.987$\\
        First MT End &  2191 & $8.007$\\
        Second MT Start &  4450 & $9.752$\\
        Second MT End &  5400 & $9.779$\\ \hline
    \end{tabular}\\
    \label{modnum}
\end{table}
The first mass-transfer stage is on the left in Fig. \ref{MT} and the second on the right. The first is much more intense, with mass-transfer rates reaching $10^{-4}\,\mathrm{M}_\odot\,\mathrm{yr}^{-1}$ during the fast portion. This contributes to the multiple solar masses transferred between the two stars during this stage. The second mass-transfer stage is much less intense and also much more erratic in terms of mass-transfer rate, because of the substantially larger initial orbital separation at this stage, the continued expansion of the orbit under mass transfer, and the fluctuations in the primary during its expansion to a helium giant. \par
The masses of the two stars, the period of the system, and the current surface hydrogen mass fraction of the primary are quantities which we can observe with high precision (see \citealt{G2023} for more details). So we wish to compare our models with the observations for these quantities to test its validity. The masses of the two stars are plotted in Fig. \ref{Mass}. We see from the plot that both final masses agree with the data presented in Table \ref{tab:upsag}. This is encouraging because it implies that the rotation of the secondary does not prevent accretion from taking place.\par
We plot the period in Fig. \ref{Per} to compare with observations. The period behaves as expected and agrees well with observations of the system in finishing at about $138\,\rm{d}$. The initial period required to match observations in these models is slightly higher than that found by \cite{G2023}. This is due to the introduction of the minimum mass-transfer $\beta$ in these models. This continually expels small amounts of mass from the system, shrinking the orbit from what it would be in the fully conservative case. Obtaining an initial period--final period relation for this type of system is nontrivial because a change in initial period also affects the timing and intensity of the mass-transfer stages. This affects the evolution of the period. \par
In Fig. \ref{SurfH} we plot the surface hydrogen mass fraction of the primary. At the end of the first mass-transfer stage most of the hydrogen in the envelope has been depleted but the system cannot yet be classed as hydrogen-deficient. This plot demonstrates the importance of the second mass-transfer stage to the production of a hydrogen-deficient binary, because this stage removes the last vestiges of hydrogen from the envelope, leading to a final mass fraction of about $10^{-3}$. This is in good agreement with the observations, and is encouraging in that we have succeeded in modelling a true hydrogen-deficient binary.
\subsection{Spin evolution}
\begin{figure*}
    \includegraphics[width=\textwidth, alt={The rotation rate with respect to critical is shown on the left. We see that it never reaches above 0.6 even during the first mass-transfer stage, and thus never hits critical. The rotational velocity then shows that during the second mass-transfer stage the peak lies at approximately 250 $km\,s^{-1}$, which matches with our observations.}]{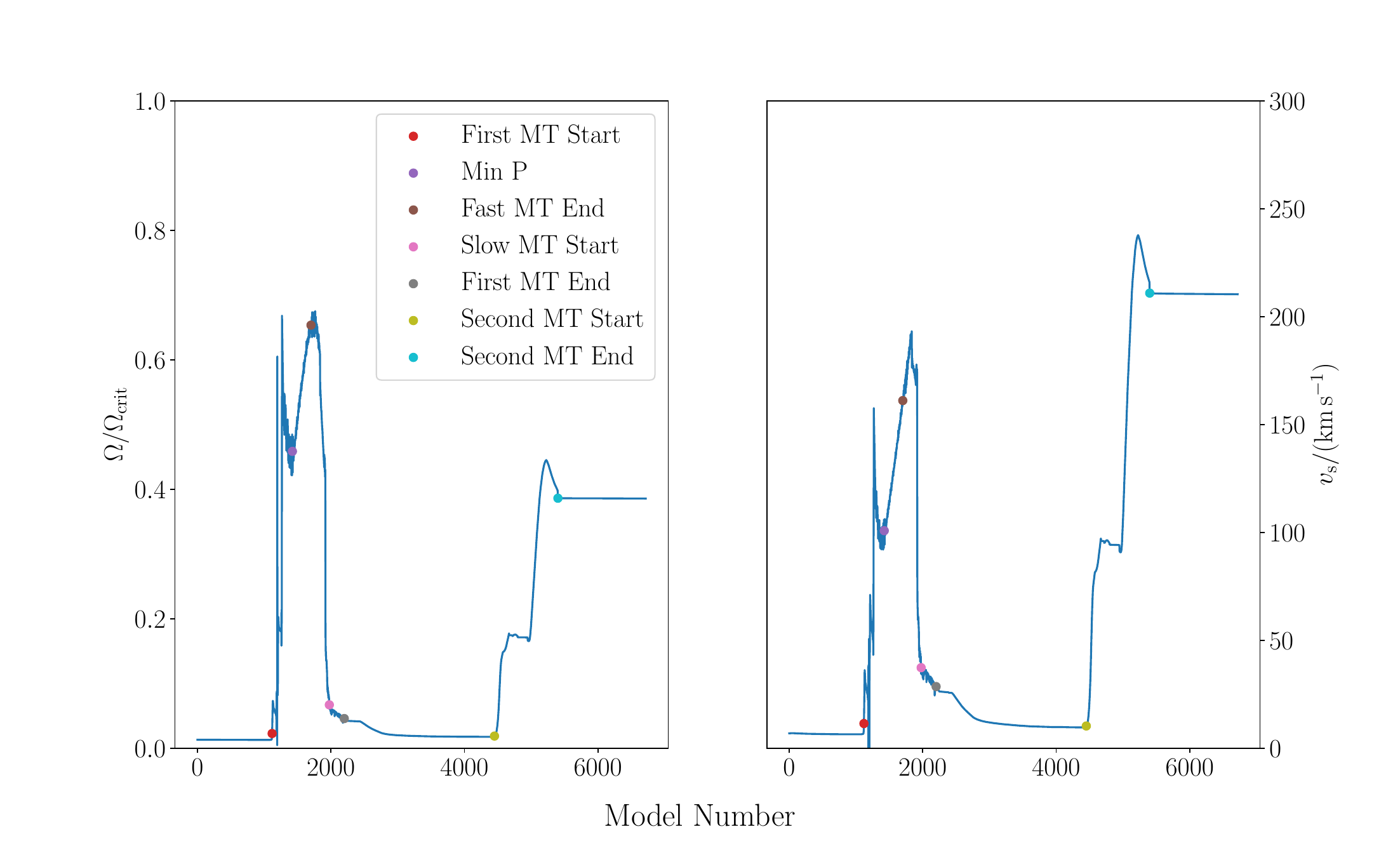}
    \vspace{-.5cm}
    \caption{Surface angular velocity $\Omega$ with respect to $\Omega_\mathrm{crit}$ (left) and equatorial rotational velocity $v_\mathrm{s}$(right) of the secondary during the evolution. The rotation rate spikes during the fast part of the first mass-transfer stage, as expected, but the magnetically coupled winds manage to keep it significantly below critical. It then drops to less than 0.1 in the transition period between fast and slow mass transfer, before increasing again to approximately 0.5 in the second mass-transfer stage. This corresponds to a surface rotational velocity of approximately $250 \,\mathrm{km\,s^{-1}}$, which agrees with the current observations of $\upsilon$ Sgr.}
    \label{rot}
\end{figure*}
\begin{figure}
    \includegraphics[width=8.5cm, alt={This plot shows the evolution of the surface magnetic field of the accreting star with model number. We see a spike in the field strength at approximately $10^5$ Gauss during the fast mass-transfer stage, followed by a period of instability, and a drop off when no mass transfer is occurring. It increases again during the second stage, but much less since the mass-transfer rate is much more subdued.}]{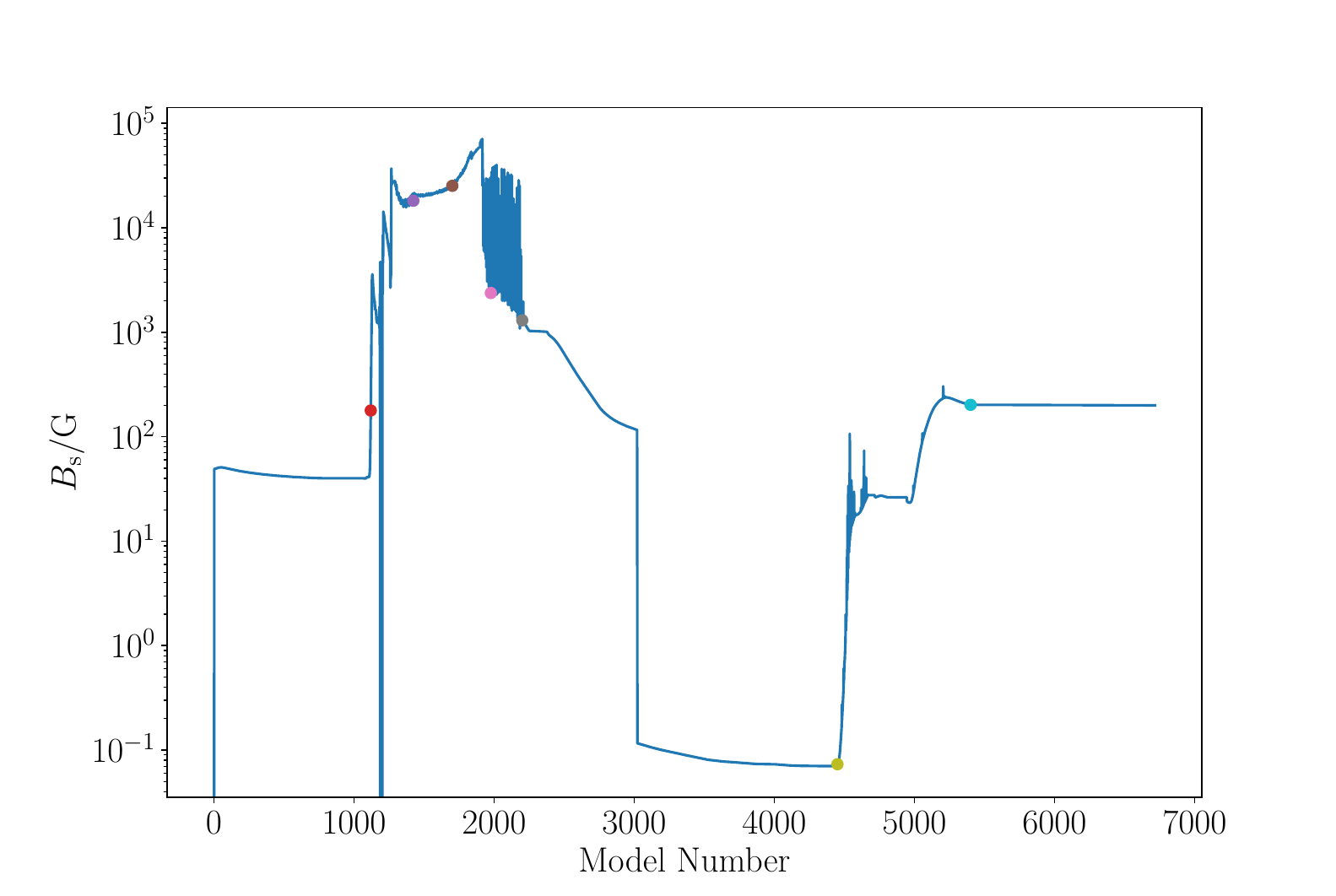}
    \vspace{-.3cm}
    \caption{The evolution of the surface magnetic field $B_\mathrm{s}$ of the accreting star with model number. This two-peaked distribution is what we would expect considering that the magnetic field should increase markedly when mass transfer is driving stronger differential rotation in the outer layers. The height of the spikes depends on the magnetic field prescription used as well as any scale factors introduced. The marked evolutionary points are as in the other figures.}
    \label{bsurf}
\end{figure}
An important focus of this investigation is the rotational evolution of the accreting star under significant mass transfer, and whether the magnetically coupled wind manages to keep the rotation below critical while also matching the observations of $\upsilon$ Sgr. In Fig. \ref{rot} we plot the stellar rotation rate with respect to critical as well as the surface rotational velocity.\par
The rotation rate does stay below critical thanks to the magnetic winds, meaning mass transfer is never halted by stellar rotation. As expected, during the fast mass-transfer stage the rotation rate does increase to above 0.6 of critical, when the mass-transfer rate is very high but, during the transition period from the fast to the slow mass transfer, $\Omega/\Omega_\mathrm{crit}$ drops to about $0.05$. This is because first the mass-transfer rate is dropping by orders of magnitude in this transition period, meaning there is significantly less spin-up of the star, so that the equilibrium rotation rate, at which the magnetic field balances the accretion, decreases in turn. Secondly the secondary is shrinking in size at this point owing to the lower accretion rate, so that $\Omega_\mathrm{crit}$ is increasing too. This precipitous drop in $\Omega/\Omega_\mathrm{crit}$ between fast and slow mass transfer persists regardless of initial conditions, initial rotation rate, and any scale factor present in the magnetic field. This also allows these models to agree with observations of rotation rates in Algol-type binaries, in which the spin is less than about $0.5$ of critical \citep{vH1990}. All of these observations were made during slow mass-transfer stages. The rotation rate then climbs once more during the second mass-transfer stage, though the height of the peak is determined by the scale factor introduced to the magnetic field during this part of the evolution (see Appendix \ref{bscaleapp}). This scale factor was modified specifically to have the surface rotation rate reach about $250\,\mathrm{km\,s^{-1}}$, to fit the observations of $\upsilon$ Sgr. \par
The magnetic field strength is shown in Fig. \ref{bsurf}. It peaks at $6\,\times\,10^4\,\rm{G}$ during the fast mass-transfer stage, before a period of rapid fluctuation between extremes approximately an order of magnitude apart during the slow mass-transfer stage. This is a numerical issue generated by the secondary fluctuating in rotational speed on either side of an equilibrium point. The strength of about $10^4\, \rm{G}$ is difficult to compare with observations because there are no comprehensive studies of the magnetic fields of accreting stars in Algol-type systems. Comparing with the magnetic fields of Ap and Bp stars, as a benchmark, indicates that these magnetic fields are higher than one would expect to see but by less than an order of magnitude \citep{A2007}. There is also, as discussed before, uncertainty in our understanding of radiatively-driven dynamos which could account for this disagreement. Artificially scaling the field during the first mass-transfer stage down by a factor of $10$ still allows the rotation rate to remain below critical during the fast mass-transfer stage. There are also no observations of magnetic fields in hydrogen-deficient binary systems to which we can compare the model magnetic field strength during the second mass-transfer stage, but strengths of around $300\,\rm{G}$ are reasonable for stellar magnetic fields (see again \citealt{A2007}). The strength of the magnetic wind itself and its modifications due to the magnetic dead zone prescription are discussed in Appendix \ref{Jdotapp}.

\section{Conclusions}
We have developed a model for rotating binary star systems undergoing mass transfer with magnetically coupled winds and star-disc coupling which allow the accreting star to successfully accrete multiple solar masses without spinning up to critical rotation. With these models, we arrive at the following conclusions.
\begin{enumerate}[labelwidth=-3\parindent]
    \item Magnetically coupled winds powered by the TSF dynamo are strong enough to allow the star to accrete multiple solar masses as observed without spinning up to critical rotation, even at high mass-transfer rates of order $10^{-4}\,\mathrm{M}_\odot\,\mathrm{yr^{-1}}$. 
    \item The accepted model of how a hydrogen-deficient binary system is formed and its general evolutionary process is well-constrained, even when rotation is added. An enforced slight non-conservation of mass transfer requires a higher initial period for $\upsilon$ Sgr than previously thought, but this is the only change required to the initial conditions.
    \item The magnetic field must be scaled down artificially during the second mass-transfer stage in order to allow the star to spin up to observed levels. This indicates a gap in our understanding of radiatively driven dynamos under accretion-driven differential rotation. This could be a result of how the rotational shear in the outer layers varies with mass-transfer rate, another poorly constrained quantity that depends on the strength of the mixing processes in said layers.
    \item In these systems, where the accretion disc is fed by mass stripped from the primary and truncated at the inner edge by a magnetic field and the outer edge by the tides of the primary, the disc is never massive enough to have a comparable moment of inertia to the star, and so cannot exert a significant torque.
\end{enumerate}
These hydrogen-deficient binary systems are ideal laboratories in which to study binary and accretion physics over orders of magnitude differences in mass-transfer rate. Furthering our understanding of how they evolve both generally and rotationally will help us to better understand the evolution and properties of interacting binary stars in general.
\section*{Acknowledgements}
We thank the anonymous referee for a constructive review of the paper. DAB acknowledges support from the David and Claudia Harding Foundation for their PhD funding. CAT thanks Churchill College for his fellowship.

%%%%%%%%%%%%%%%%%%%%%%%%%%%%%%%%%%%%%%%%%%%%%%%%%%
\section*{Data Availability}
The input files necessary to reproduce our simulations and the data products of the specific simulation referred to in this paper are available at \href{https://doi.org/10.5281/zenodo.17046074}{https://doi.org/10.5281/zenodo.17046074}.



\bibliographystyle{mnras}
\bibliography{bib} % if your bibtex file is called example.bib

%\begin{thebibliography}{99}
%\end{thebibliography}

%%%%%%%%%%%%%%%%%%%%%%%%%%%%%%%%%%%%%%%%%%%%%%%%%%

%%%%%%%%%%%%%%%%% APPENDICES %%%%%%%%%%%%%%%%%%%%%

\appendix

\section{Magnetic Field Scalings}\label{bscaleapp}

\begin{figure*}
    \includegraphics[width=\textwidth, alt={The variance of $\Omega$ and the surface rotation rate of the accretor with respect to model number for different scale factors on the magnetic field, from 1 to $10^{-3}$. The spacing is not uniform, but does follow the trend of smaller scale factors leading to larger rotation rate, as one would expect.}]{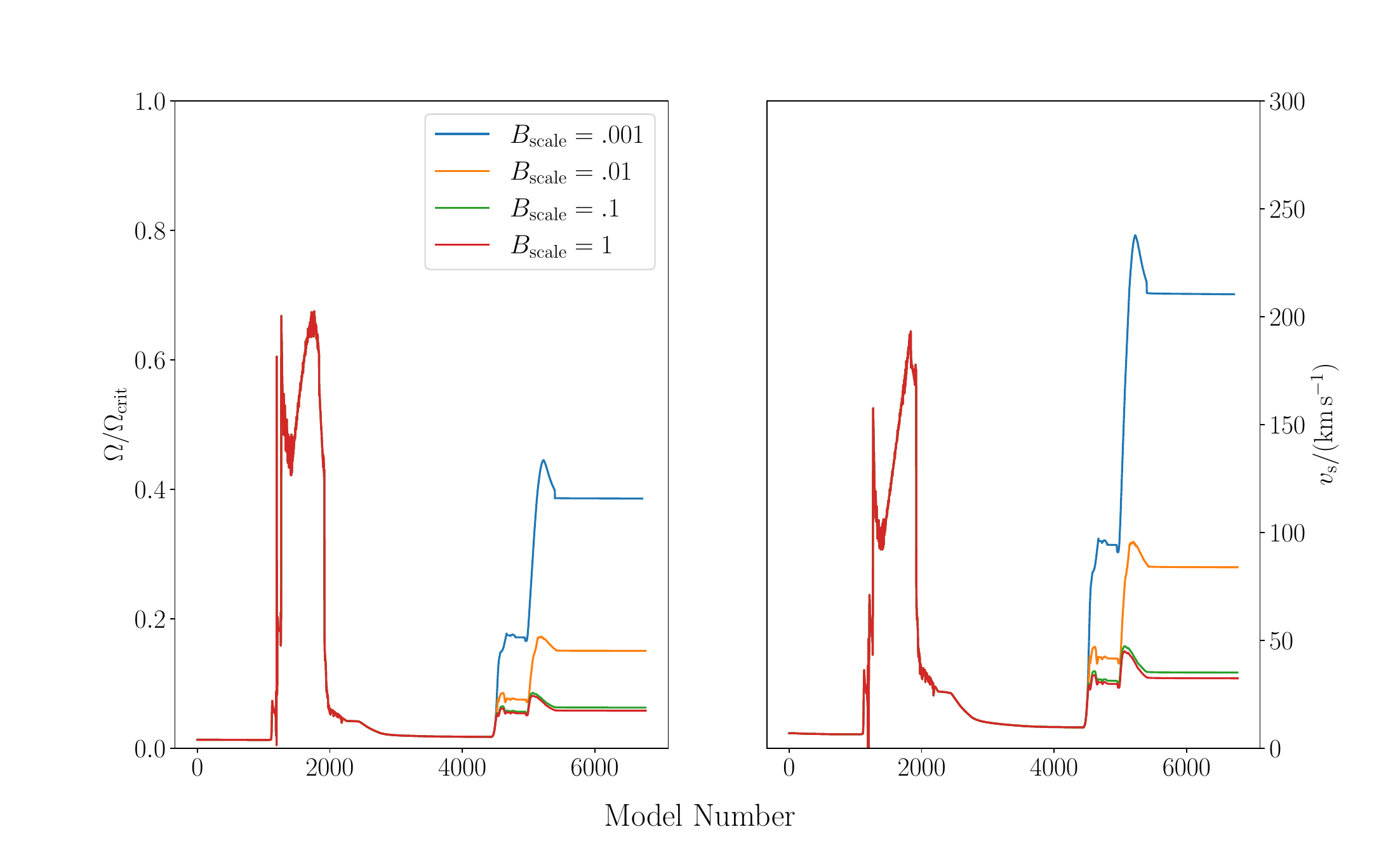}
    \vspace{-.3cm}
    \caption{The left panel shows the variability of $\Omega/\Omega_\mathrm{crit}$ with model number for different artificial scalings introduced into the magnetic field prescription, while the right panel shows the same for the surface velocity. The overall trend is that a smaller scale factor leads to more rapid rotation. This is expected because it reduces the strength of the magnetic wind. }
    \label{bscaleomega}
\end{figure*}

\begin{figure}
    \includegraphics[width=8.5cm, alt={We see here the magnetic field strength varying for different scale factors. Three are relatively evenly spaced, while the fourth is much closer to its neighbour.}]{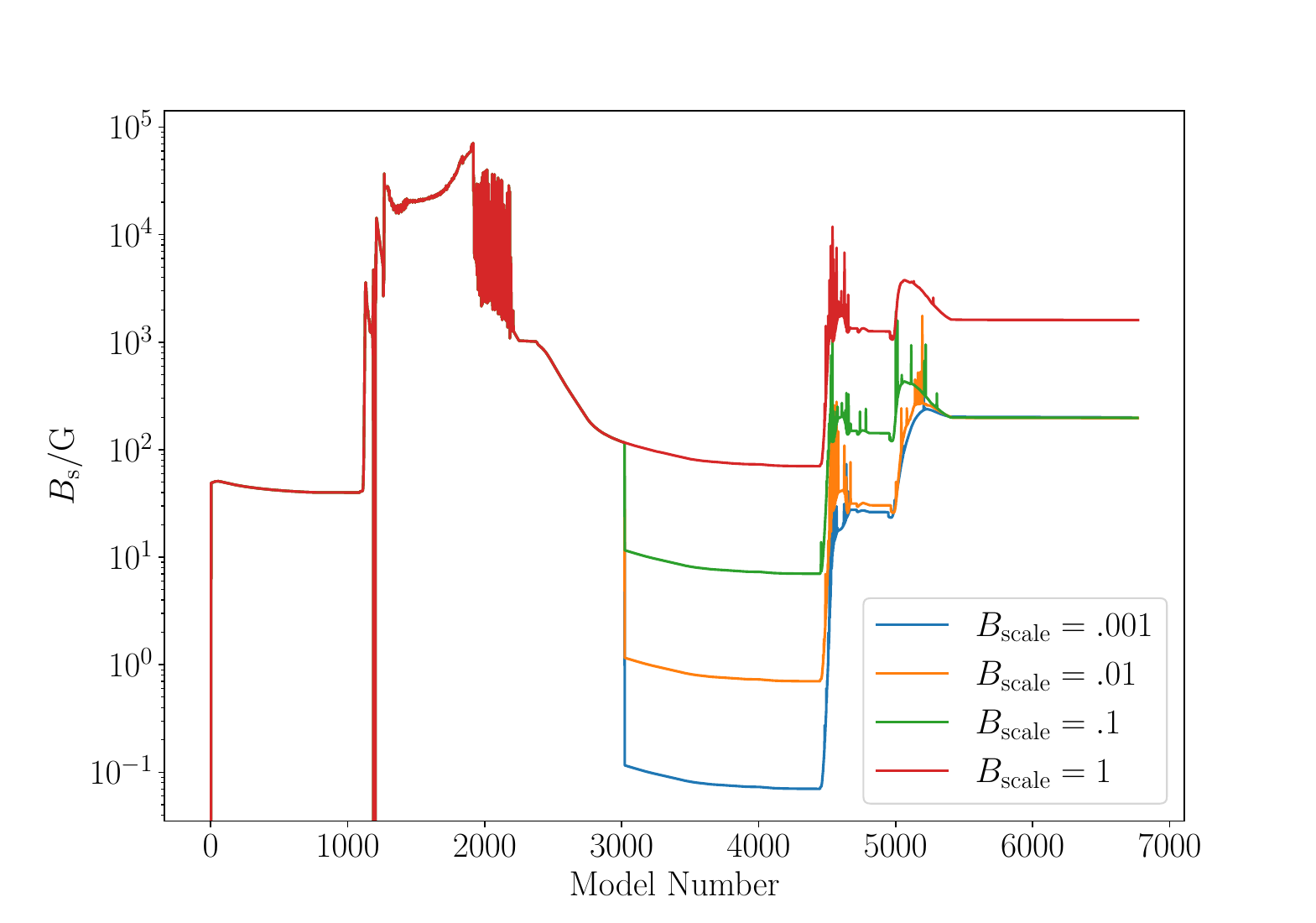}
    \vspace{-.3cm}
    \caption{Magnetic field strength for different scale factors. These are not spaced uniformly as one might expect because the magnetic field strength also depends on the spin. This is what causes the very small difference in magnetic field strength between scale factors of $10^{-2}$ and $10^{-3}$.}
    \label{bscalebsurf}
\end{figure}

Figs. \ref{bscaleomega} and \ref{bscalebsurf} show how the angular velocities and the magnetic field strength vary with the artificial scaling factor introduced for the second mass-transfer stage. The shape of the curves of both $\Omega/\Omega_\mathrm{crit}$ and $v_\mathrm{s}$ with respect to model number do not change significantly with the scaling. The only change is where the equilibrium point is established between the spin-down of the magnetic winds, the strength of which increases with $\Omega$, and the spin-up by accretion. The difference in equilibrium between scaling factors is not uniform because the dependence of the magnetic winds on $\Omega$ is not linear. Similarly, the magnetic field strength does not scale by uniform factors of 10 because the equilibrium strength is also determined by the mass-transfer rate and stellar spin. Interestingly the difference in strength of the magnetic field between scale factors of $10^{-2}$ and $10^{-3}$ during the second mass-transfer stage is very small, indicating that the change in rotation rate is enough to make up the factor of 10 taken away from the field.

\section{Angular Momentum Losses}\label{Jdotapp}
\begin{figure}
    \includegraphics[width=8.5cm, alt={The strength of the angular momentum losses due to the magnetically coupled wind are plotted with respect to model number. There are two peaks corresponding to the two mass-transfer stages, with the first higher by multiple orders of magnitude as expected.}]{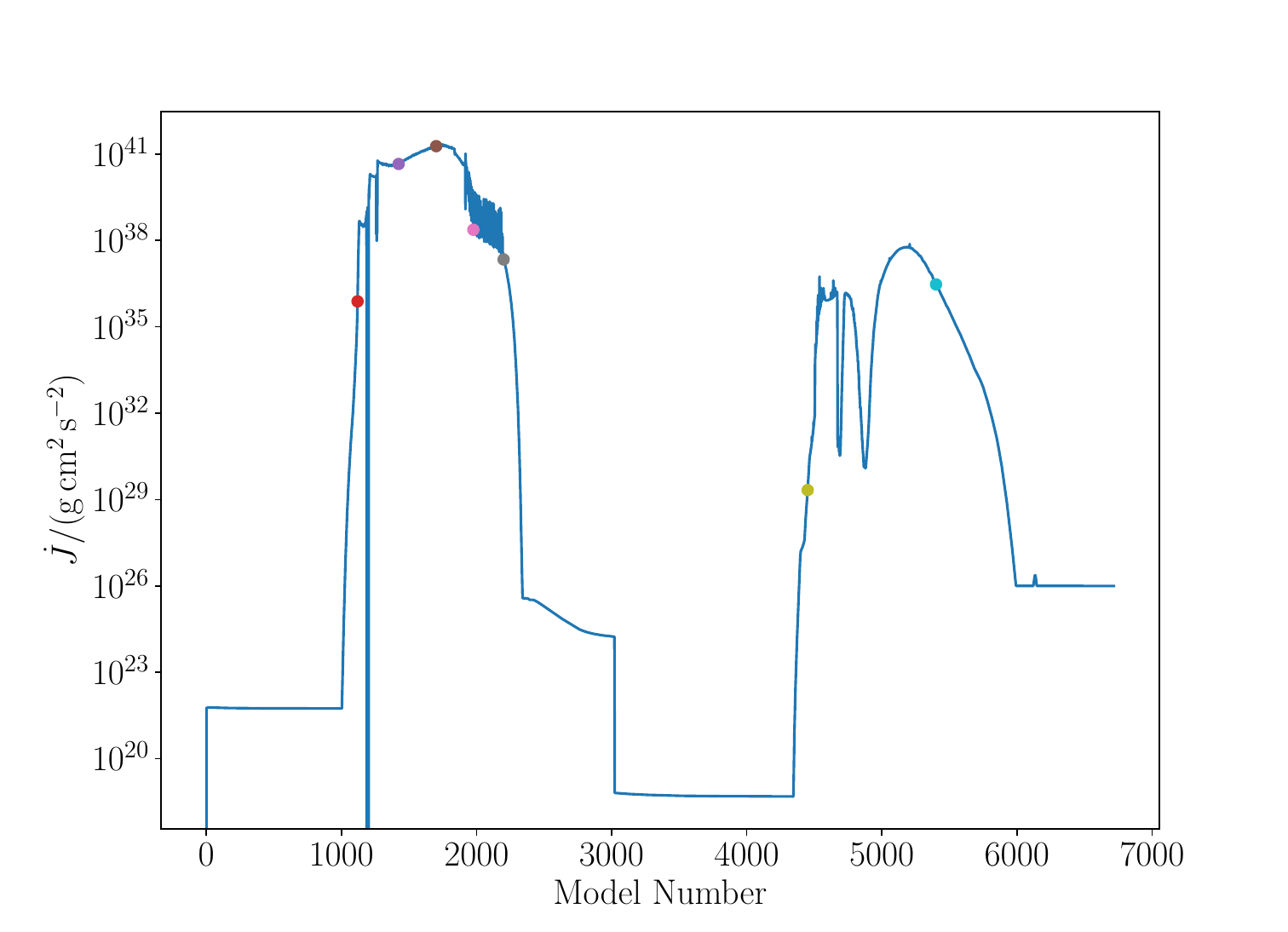}
    \vspace{-.3cm}
    \caption{Strength of the magnetically coupled wind in terms of the total angular momentum it removes from the star. As expected, this is at its maximum during the first mass-transfer stage when the amount of material expelled from the system and the magnetic field are at their highest. The second region of activity corresponds to the second mass-transfer stage.}
    \label{jdot}
\end{figure}
\begin{figure}
    \includegraphics[width=8.5cm, alt={}]{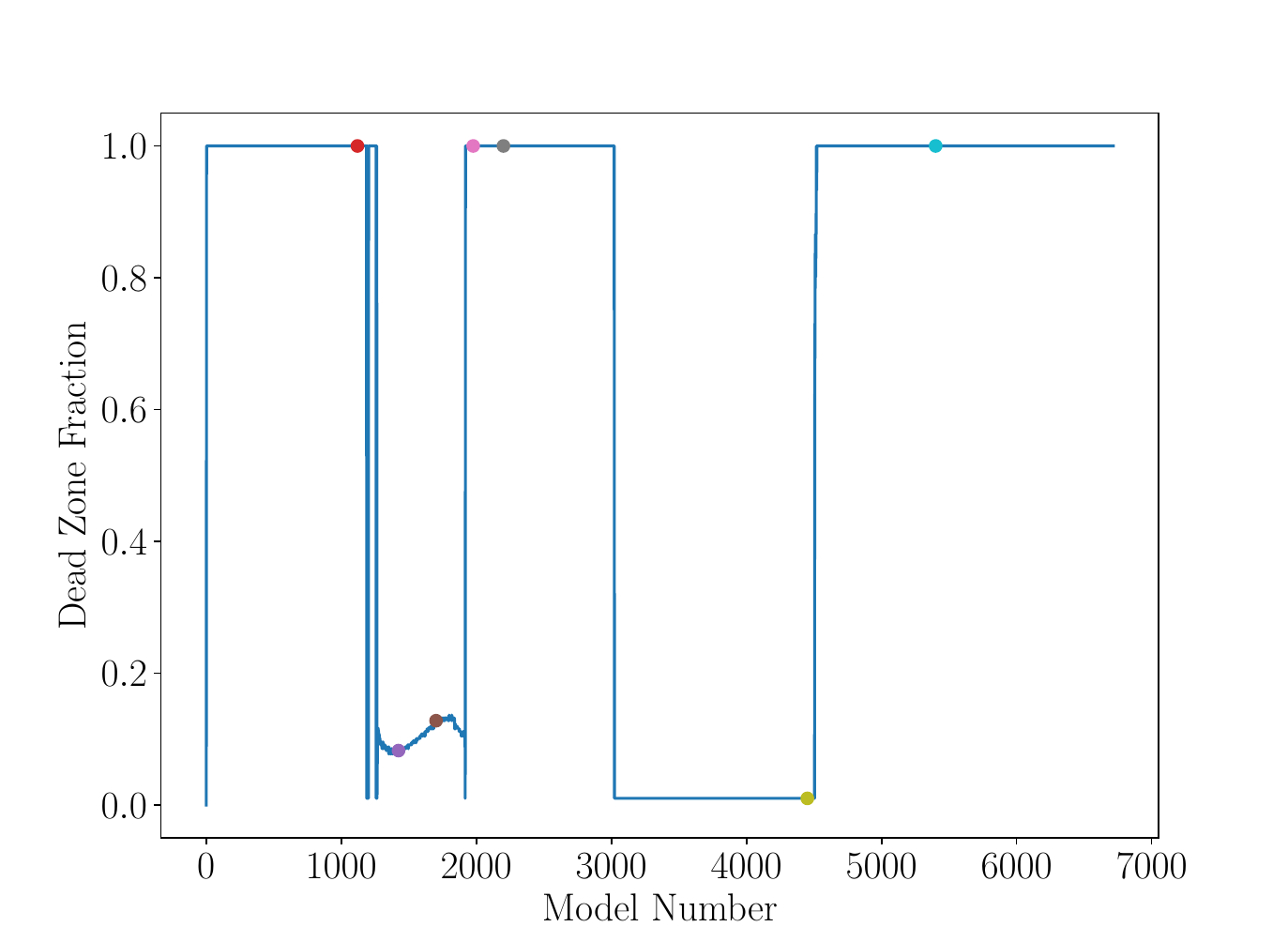}
    \vspace{-.3cm}
    \caption{Multiplicative factor applied to $\dot{J}_\mathrm{wind}$ to account for magnetic dead zones. This only has an effect during the fast mass-transfer stage. It is important during this period, reducing the strength of the magnetic wind by an order of magnitude. The period between models 3000 and 4500 is in between the two mass-transfer stages and so the dead zone fraction is set to 0 by default.}
    \label{deadzone}
\end{figure}
Fig. \ref{jdot} shows the strength of the angular momentum losses in magnetic winds. As expected, we see two peaks corresponding to the two mass-transfer stages. The losses are much more significant during fast mass transfer, because both the spin rate of the star and the amount of mass expelled from the system are large. This $\dot{J}$ also depends on the magnetic dead zone prescription, which generates a multiplicative factor to be applied to $\dot{J}$ in order to account for dead zones around the star. The magnitude of this factor is plotted in Fig. \ref{deadzone}. The figure demonstrates that dead zones only have an effect on the star for a brief period during the fast mass-transfer stage but, during this stage, the effect is non-negligible, reducing $\dot{J}$ by approximately an order of magnitude. This allows the star to spin up more than it otherwise would have been able to but, importantly, the angular momentum losses are still sufficient to keep it below critical.

%%%%%%%%%%%%%%%%%%%%%%%%%%%%%%%%%%%%%%%%%%%%%%%%%%

% Don't change these lines
\bsp	% typesetting comment
\label{lastpage}
\end{document}